\documentclass{article}

\usepackage{arxiv}

\usepackage[utf8]{inputenc} 
\usepackage[T1]{fontenc}    
\usepackage{hyperref}       
\usepackage{url}            
\usepackage{booktabs}       
\usepackage{amsfonts}       
\usepackage{nicefrac}       
\usepackage{microtype}      
\usepackage{lipsum}
\usepackage{graphicx}
\graphicspath{ {./images/} }

\usepackage{subcaption} 
\usepackage{amsmath} 
\usepackage{algorithm}
\usepackage{algpseudocode} 
\usepackage{xcolor} 
\usepackage{listings} 
\usepackage{array}      
\usepackage{caption}    
\usepackage{cite} 
\title{Iterative Optimization Algorithm with Shape Prior to Reduce Noise and Artifacts for Photoacoustic Imaging}

\author{
 Zhang Yu\\
  Department of Biomedical Engineering\\
  College of Future Technology \\
  Peking University\\
  \texttt{zyuaiyi1\_@stu.pku.edu.cn} \\
   \And
 Li Shuang \\
  Department of Biomedical Engineering\\
  College of Future Technology \\
  Peking University\\
  \texttt{jaeger\_ls@stu.pku.edu.cn} \\
   \And
 Wang Yibing \\
  Department of Biomedical Engineering\\
  College of Future Technology \\
  Peking University\\
  \texttt{ddffwyb@pku.edu.cn} \\
   \And
Sun Yu \\
  Department of Biomedical Engineering\\
  College of Future Technology \\
  Peking University\\
  \texttt{sunyu@stu.pku.edu.cn} \\
   \And
Xiang Wenyi \\
  Electronics\\
  School of Electrical and Electronic Engineering\\
  Nanyang Technological University \\
  \texttt{WENYI004@e.ntu.edu.sg} \\
  \And
Li Changhui \\
  Department of Biomedical Engineering\\
  College of Future Technology \\
  Peking University\\
  \texttt{chli@pku.edu.cn} \\
  \AND
}

\begin{document}
\maketitle
\date{}
\begin{abstract}
\hspace*{2em}Although exact reconstruction algorithms exit for photoacoustic imaging (PAI), in reality its image quality is inevitably influences by system noise and artifacts. In this study, we propose an iterative optimization method for PAI reconstruction, called the regularized iteration method with shape prior (RISP). The shape prior is a probability matrix derived from multiple reconstruction results via randomly partial array elements using any reconstruction algorithms, such as widely used Delay-and-Sum (DAS) and Filtered Back-Projection (FBP). In the probability matrix, high-probability values indicate high coherency among multiple reconstruction results at those positions, suggesting a high likelihood of representing the true imaging results. In contrast, low-probability values indicate higher randomness, leaning more towards noise or artifacts. As a shape prior, this probability matrix, together with the orginal PAI result using all array elements, guides the iteration and regularization of the PAI results. The probability matrix is taken as a prerequisite for improving the original reconstruction results, and the optimizer is used to further iterate the imaging results to remove noise and artifacts and improve the imaging fidelity. The simulation and real PAI animal and human results demonstrated our method can substantially reduce both noise and artefacts without the need of hardware upgrade.
\end{abstract}

\newpage

\section{Introduction}
\hspace*{2em}PAI has emerged as a promising imaging modality that combines the high spatial resolution of ultrasound with the rich optical contrast of biological tissues\cite{1}. This hybrid imaging technique has found wide-ranging applications in biomedical imaging due to its ability to provide detailed structural and functional information, particularly in scenarios such as vascular imaging\cite{2}, tumor detection\cite{3}, and brain imaging\cite{4}. However, in reality, the image quality of PAI generally suffers from inherent limitations that can degrade the quality of reconstructed results, such as noise, artifacts, and incomplete data acquisition caused by sparse sampling or partial array detection. Traditional image reconstruction methods, while effective to some extent, often struggle to handle such challenges, especially when the imaging system faces conditions like a sparse view or other constraints that exacerbate result artifacts.\cite{5} These approaches may fail to fully exploit the inherent redundancies or structural priors present in the data, leading to suboptimal performance in terms of reconstruction fidelity and artifact suppression.\\
\hspace*{2em}To tackle these challenges, researchers have utilized iterative reconstruction (IR) methods to enhance result quality when using sparse view. One of the earliest IR approaches for PAI was introduced by Paltauf et al., who iteratively minimized the discrepancies between observed and simulated projections to improve image reconstruction.\cite{6} Subsequent IR techniques incorporated various physical factors related to imaging. Later, Deán-Ben et al. developed a precise model-based 3D reconstruction algorithm that used a sparse matrix model to compute theoretical pressure\cite{7}, resulting in fewer artifacts compared to back-projection methods. Wang et al.\cite{8,9} demonstrated that iterative penalized least-squares methods, constructed on discrete-to-discrete imaging models with expansion functions over different fixed spatial grids, and using either quadratic smoothing or total variation (TV) norm constraints, could significantly improve the performance of 3D PAI systems for small animals. Similarly, Huang et al.\cite{10} proposed forward and backward operators based on k-space pseudospectral methodologies. While these IR techniques have shown excellent results, While these IR techniques have shown excellent results, the computing burden in both time and memory consumption is still a serious challenge for their implementations, especially for large-scale 3D PAI.\cite{11} Researchers have, therefore, explored novel alternatives. Arridge et al. addressed this by introducing efficient numerical implementations of adjoint operators for PAI reconstruction and applying compressed sensing via Bregman iteration to reduce sensor requirements.\cite{12}  However, their method was computationally expensive and restricted to voxel counts on the scale of $10^{6}$. Shang et al.\cite{13} took a different approach by building a forward model using directly measured graphite point sources, but the inefficiencies in optimization and inaccuracies in the model itself limited its practicality. Deep learning-based methods for PAI reconstruction improve computational efficiency and image quality\cite{14,15,16}  but depend on large, specialized datasets that are hard to obtain and struggle to generalize across different imaging systems or conditions, limiting their clinical applicability.\\
\hspace*{2em}Unlike the traditional IR algorithms that demands repeating solving forward model for every iteration circle, here, we proposed a new optimization method for the results of photoacoustic imaging reconstruction that does not need solving the forward model. The key idea of this method is that the similarity of the structure representing the true imaged object is much higher than the random noise and variation artifacts highly related to the arrangement of sensor positions and numbers. Therefore, we construct a probability matrix based on a large number of image reconstruction results from sub-sets of randomly picked original array elements using any reconstruction algorithm, such as DAS and BP. And the probability matrix is obtained through calculation methods. The different positions of the matrix reflect the probability that the position is the real object in the imaging space. As a shape prior\cite{17}, this probability matrix guides the iteration and regularization of the originally reconstructed results using total array elements. This method is not affected by specific reconstruction algorithms, such as DAS and UBP, and it can be applied for any array configurations. By leveraging the shape prior based on possibility matrix and iteratively refining reconstruction results, this method substantially improved image quality while keeping quantitative accuracy. Both simulation and experimental results demonstrate the effectiveness of the proposed method in reducing noise and artifacts under sparse-view conditions while preserving critical structural information in PAI reconstructions.\\ 
\hspace*{2em}This paper is organized as follows. In Section 2, we present the theoretical framework of the proposed method, detailing the construction of the probability matrix, the integration of the shape prior, and the iterative algorithm employed. Section 3 describes the simulation and experimental results, demonstrating the advantages of our method. Finally, Section 4 concludes the paper by discussing the implications of the proposed method and exploring potential future developments for RISP in computational imaging applications.

\newpage
\section{Method}
\label{sec:headings}
\hspace*{2em}This part is mainly divided into two parts. The first part is the construction of probability matrix generated by random partial element signals, which provides a shape prior knowledge for subsequent regular iteration; The second part is the specific iteration optimization scheme.

\subsection{Construction of Probability Matrix}

\subsubsection{Reconstruction of Random Selected Element Signals}
\hspace*{2em}For a PAI array system with the element number $N$ and temporal sampling number $M$, the data dimension of recorded signal is $N\times M$. Besides using all $N$ elements to do image reconstruction, we randomly select $x$ ($x<N$) sub-elements for reconstruction. The total number of selection is A (A = $c_N^x$), and we only randomly pick k results ($k<A$), naming $R_1, R_2,...,R_k$. (The choices of $x$ and $k$ are described in detail in subsequent paragraphs.)
\subsubsection{Construction of Probability Matrix}

\hspace*{2em}In 3D PAI, the probability matrix is defined as $P(x,y,z)$.
\begin{align}
  P(x,y,z)=D(x,y,z)\odot\int_{v_{\min}}^{v_{\max}}\Phi(v|\mu,\sigma)\odot\mathrm{u}(D(x,y,z)-v)\mathrm {d}v 
\end{align}
\hspace*{2em}where, $\mathrm{u}(x)$ is the unit step function, and $D(x,y,z)$ is calculated by
\begin{equation}
  D(x,y,z)=\frac{(\frac{1}{k}(R_1+R_2+\cdots+R_k))^{2}}{\frac{1}{k}(R_1^{2}+R_2^{2}+\cdots+R_k^{2})}=\frac{(R_1+R_2+\cdots+R_k)^{2}}{k(R_1^{2}+R_2^{2}+\cdots+R_k^{2})}
\end{equation}
\hspace*{2em} $v_{min}=\mathrm{min}(D(x,y,z)),\ v_{max}=\mathrm{max}(D(x,y,z)),\ \mu=(v_{max}+v_{min})/2,\ \sigma = (v_{max}-v_{min})/6$, and $\Phi(v|\mu,\sigma)$ is the cumulative distribution function of the normal distribution.
\begin{equation}
\Phi(v|\mu,\sigma)=\frac{1}{\sigma\sqrt{2\pi}}\int_{-\infty}^{v}e^{\frac{-(t-\mu)^2}{2\sigma^2}}\mathrm{d}t
\end{equation}
\hspace*{2em}In order to have a more intuitive physical meaning for the subsequent regular iteration process,  we use normalized $P_{norm}$ $(0<P_{norm}<1)$.
\begin{equation}
{P_{norm}=\frac{P-\mathrm{min}(P)}{\mathrm{max}(P)-\mathrm{min}(P)}}
\end{equation}
\hspace*{2em}The Matrix $P_{norm}(x,y,z)$ is a probabilistic representation of spatial data distributions. The function integrates the local property $D(x,y,z)$, which quantifies the concentration or uniformity of measured attributes $R_1,R_2,...,R_k$ at a given spatial point $(x,y,z)$, defined as the ratio of the squared arithmetic mean to the mean of squared values. This makes $D(x,y,z)$ a measure of localized data consistency, where higher values indicate greater consistency. The introduction of the Gaussian cumulative distribution-based weighting factor $\Phi(v|\mu,\sigma)$ combines local signal intensities with global distribution characteristics, ensuring that the weight assignment is both scientifically robust and consistent. The Gaussian distribution, determined by its mean $\mu$ and standard deviation $\sigma$, dynamically reflects the global distribution pattern of signals. Strong signal regions above the mean receive higher weights, while weaker signals are appropriately suppressed but not entirely discarded. This smoothing property avoids the adverse effects of outliers or noisy signals on the results while presenting a clear and hierarchical structure of signal importance across the global range. Consequently, this weighting approach enhances signal saliency, suppresses background noise, and selectively preserves significant regions, achieving more robust and efficient reconstruction. This integral effectively blends the local signal characteristics with global weighting, enabling the assignment of weights that emphasize real signals while preserving the overall smooth and consistent distribution across the signal space.

\subsection{Regularized Iteration Optimization}
\hspace*{2em}In this section, we will use the probability matrix $P_{norm}$ obtained in the previous section as a shape prior and design a suitable loss function for regular iterative optimization of the original reconstruction results of the entire array signals.

\subsubsection{Shape Prior and Loss Function}
\paragraph{Shape Prior}
The feasibility of using $P_{norm}$ as a shape prior lies in the fact that it directly provides the probability distribution information of each voxel or pixel region, which can be used to indicate the model's tendency to strengthen certain regions (high probability regions) in the reconstruction results and its inhibition demand for other regions (low probability regions). This prior information can effectively guide the optimization process, making the result more in line with the expected structural characteristics, while reducing the impact of noise or artifacts. The rational constraint of shape information is realized mathematically, and the optimization stability of shape consistency is improved. It has clear feasibility and physical significance.

\paragraph{Loss Function}
The result of reconstructing all array signals using the algorithm to be optimized is $R_N$. The result of each iteration optimization is $R_{op}$. To enhance the generalizability of the algorithm and reduce the need for extensive parameter adjustments during application, $R_N$ and $R_{op}$ are normalized to the range of $[0,1]$ after taking the absolute value operation. The loss function consists of two terms: data consistency loss term\cite{18} and regularization loss term\cite{19}.

\subparagraph{Data Consistency Loss Term}
The data consistency loss term is represented by $Loss_{dc}$.
\begin{equation}
{Loss_{dc}}=||R_N - R_{op}||^2_{F}
\end{equation}
Its corresponding gradient is $gradient_{dc}$. 
\begin{equation}
{gradient_{dc}=\frac{\partial{Loss}_{dc}}{\partial R_{op} }=2\odot(R_N - R_{op})}
\end{equation}
\hspace*{2em}The role of the data consistency loss term is to ensure that the optimized results do not deviate too much from the original reconstruction results, thereby maintaining physical consistency and quantitative accuracy. Although $R_N$ contains noise or artifacts, it is still close to the true image. Due to the regularization introduced in the optimization process, the $R_op$ of the optimization result may deviate from the original data, and even lose the important features of original signals. Therefore, the data consistency loss term imposes constraints in the optimization process, forcing the optimization results to be consistent with the original reconstructed data as much as possible, so that the optimized volume data can gradually de-noise and enhance the structure while still reflecting the main features and information of the input data. This is a balancing mechanism that avoids distortion caused by excessive regularization and strengthens the physical confidence and data integrity of the optimized results, which is essential for quantitative accuracy.
\subparagraph{Regularization Loss Term}
The regularization loss term is represented by $Loss_{rg}$. 
\begin{equation}
{Loss_{rg}}=||(1-P_N)\odot R_{op}||^2_{F}
\end{equation}
Its corresponding gradient is $gradient_{rg}$. 
\begin{equation}
{gradient_{rg}=\frac{\partial{Loss}_{rg}}{\partial R_{op} }=2\times (1-P_N)\odot R_{op}}
\end{equation}
\hspace*{2em}The significance of regularization loss term is to use the shape prior knowledge to constrain the optimization process. As we mentioned in the previous section, in the actual reconstruction, the results obtained by using the traditional reconstruction algorithm are accompanied by noise and artifacts, which makes the reconstruction results bad and even difficult to reflect the real shape of the target. By combining the prior information provided by the prior probability distribution $P_N$, the regularization term can actively suppress the data in the low probability region, thereby reducing the interference both of artifacts and noise to the solution in the reconstruction process. The design idea of the regularization term is to guide the optimal solution to cluster in the high probability region (these regions are more in line with the prior physical knowledge), that is, the penalty of the high probability region is low, while the value of the low probability region is suppressed, that is, the penalty of the low probability region is high. This constraint can make the optimization more robust, that is, reduce overly radical updates in areas with greater uncertainty, and ultimately promote the optimized results to have both prior rationality and data authenticity, thus generating reconstruction results with higher reliability.

\subsubsection{Optimizer Selection and Regularized Iteration Process}
\paragraph{Optimizer Selection}
We selected the Adam optimizer (Adaptive Moment Estimation) as our optimization algorithm. Adam is a first-order gradient-based optimization algorithm that combines the advantages of momentum and root mean square propagation (RMSProp)\cite{20}, making the optimization process more stable and efficient in complex reconstruction tasks. Specifically, the momentum mechanism reduces oscillations and gradient noise by applying an exponentially weighted average to the gradients, while RMSProp's adaptive learning rate ensures proper adjustments of the learning rates across different variable dimensions. Additionally, Adam incorporates bias correction, which rectifies the biases during early iterations, thereby improving stability. Overall, with its fast convergence and adaptive features, the Adam optimizer meets the requirements of our task.
\paragraph{Regularized Iteration Process}The following is the specific algorithm pipeline of iterative optimization:

\begin{algorithm}[H]
\caption{}
\begin{algorithmic}[1] 
    \State \textbf{Input: } 
    \State \hspace{1.0cm} $R_N$: Reconstruction results to be optimized  $\quad$ \Comment{$[-1,1]$}
    \State \hspace{1.0cm} $P_{norm}$: Probability matrix  $\quad$ \Comment{$[0,1]$}
    \State \hspace{1.0cm} $\lambda_{con}$, $\lambda_{reg}$: Data consistency term coefficient, regularization coefficient
    \State \hspace{1.0cm} $lr$, $num\_iters$: Learning rate, Number of iterations
    \State \hspace{1.0cm} $optimizer \gets AdamOptimizer(R_N, lr=learning\_rate)$ 
    
    \State \textbf{Output: } $R_{op}$: Iteration result  $\quad$ \Comment{$[-1,1]$}
    \For{$iteration \gets 0$ \textbf{to} $num\_iters$}
        \State \textbf{Step 1: }
        \State \hspace{1.0cm} $gradient_{data\_consistency} \gets 2\times (R_N - R_{op})$
        
        \State \textbf{Step 2: }
        \State \hspace{1.0cm} $gradient_{prior\_reg} \gets 2\times (1-P_N)\odot R_{op}$

        \State \textbf{Step 3: }
        \State \hspace{1.0cm} $total\_gradient \gets \lambda_{con} \cdot gradient_{data\_consistency} + \lambda_{reg} \cdot gradient_{prior\_reg}$

        \State \textbf{Step 4: }
        \State \hspace{1.0cm} $new\_lr \gets lr $
        \State \hspace{1.0cm} optimizer.set\_lr($new\_lr$)
        
        \State \textbf{Step 5: }
        \State \hspace{1.0cm} $step \gets optimizer.update(total\_gradient)$
        \State \hspace{1.0cm} $R_{op} \gets R_{op} + step$

    \EndFor

\end{algorithmic}
\end{algorithm}

\section{Results}
\hspace*{2em}In this session, we use 3D simulation data and in vivo animal experiment data to verify the performance of RISP algorithm.
\subsection{Simulation Data Results}
\subsubsection{3D PA Image Reconstruction under 1024 Hemispherical System}
\begin{figure}[htbp]
    \centering
    \begin{minipage}[t]{0.8\textwidth} 
        \centering
    
        \begin{subfigure}[b]{\textwidth}
            \centering
            \includegraphics[height=5.2cm]{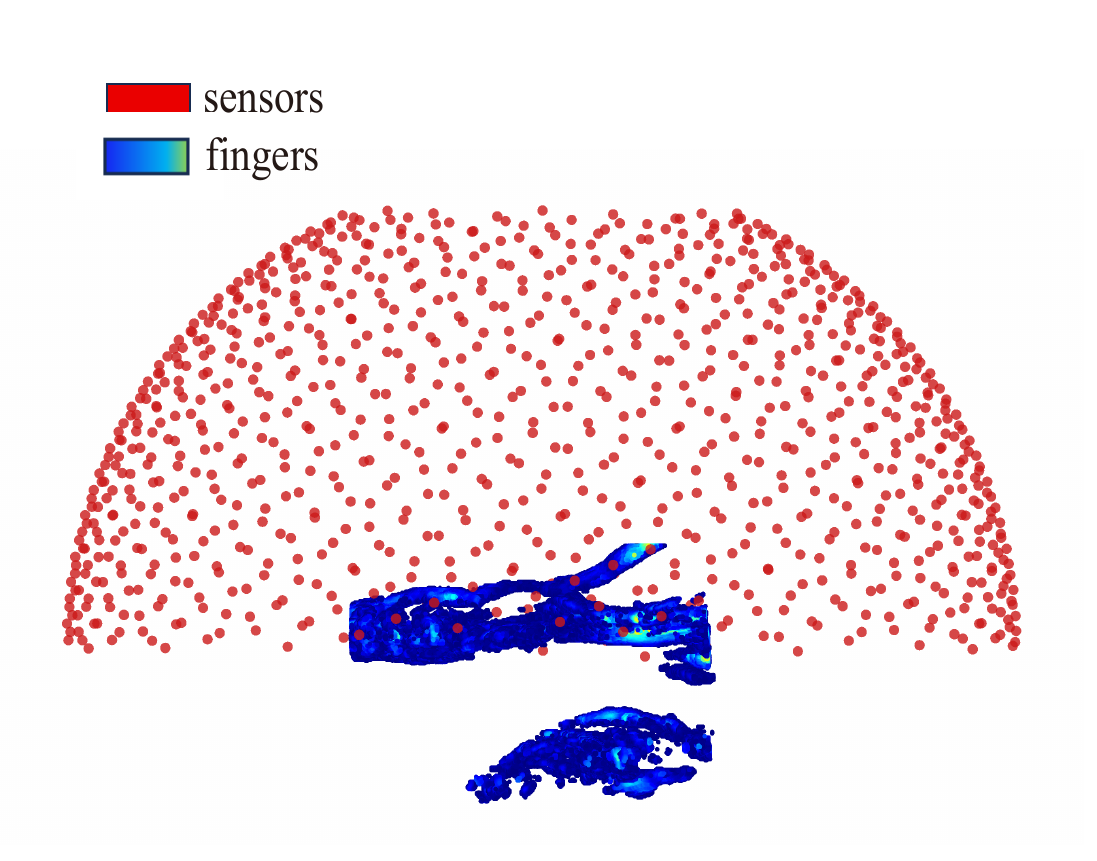} 

        \end{subfigure}
    \end{minipage}
    \caption{The position of the fingers and sensors}
    \label{fig:figure3}
\end{figure}

\hspace*{2em}We first used simulated hand vessels as the target tissue to demonstrate the high performance of the RISP algorithm. The simulated data is based on the 3D PA imaging results of human peripheral hand vessels from another work\cite{21}, which was used as the ground truth. At a grid point spacing of 0.2 mm, the total grid size of the imaging area is 128$\times$320$\times$256 (25.6 mm $\times$64.0 mm $\times$51.2 mm). In order to simulate 3D PA imaging, the radius of the hemispherical system is 4cm, and a gap with a height of 1cm is opened at the top as the laser light inlet, and the remaining positions are evenly distributed with 1024 array elements(Figure \ref{fig:figure3} is imaging setup). Then we obtained PA signals collected by these 1024 arrays through forward propagation of the K-Wave toolbox\cite{22}.  

\hspace*{2em}The above imaging system was selected for simulation to generate sparse photoacoustic signals in order to evaluate the performance of the RISP algorithm. The simulation is based on 3D photoacoustic imaging data of real hand vasculature as the "ground truth," using a high-resolution grid with a resolution of 0.2 mm (128×320×256) to capture the intricate structures of the vasculature. The system is designed as a hemispherical array with a radius of 4 cm, featuring a laser entry point at the top with a height of 1 cm to create a missing-view angle. The array is uniformly distributed with 1024 sensors to simulate a typical photoacoustic imaging device. This simulation approach closely approximates real-world scenarios, provides a controlled evaluation standard, and offers a reliable basis for validating the performance of the RISP algorithm in photoacoustic imaging.

\paragraph{Comparison of Results before and after Optimization\\}

\begin{figure}[htbp]
    \centering
    \begin{subfigure}[b]{1\textwidth} 
        \centering
       
        \includegraphics[width=1\textwidth]{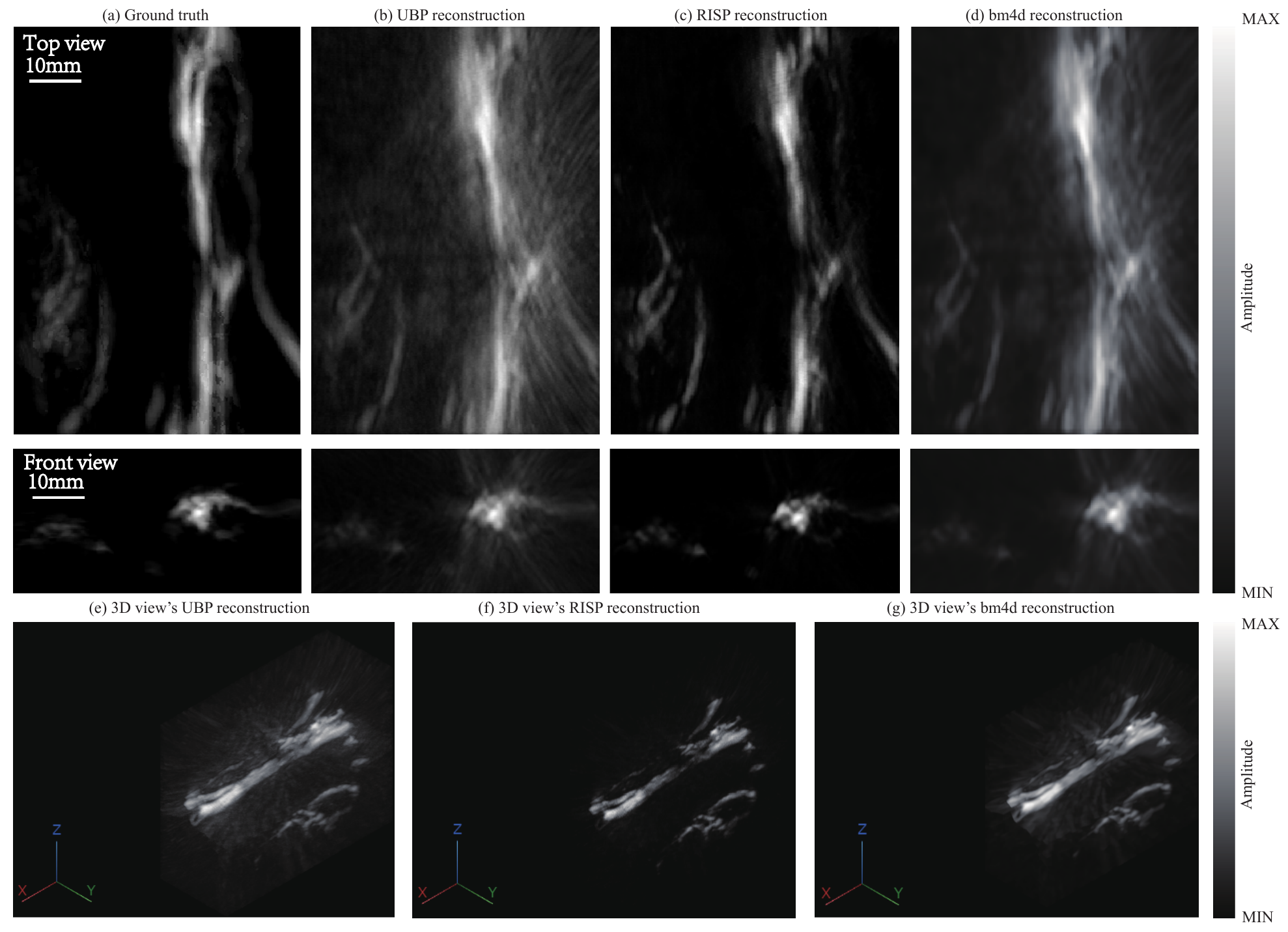} 
        \label{fig4:a}
    \end{subfigure}
    
    \caption{Comparison of 3D photoacoustic reconstruction results under 1024 hemispherical system. (a) Finger's top-view maximum amplitude projection, front-view maximum amplitude projection of the acoustic source.\\ 
    (b) Finger's top-view maximum amplitude projection, front-view maximum amplitude projection of the UBP result. \\
    (c) Finger's top-view maximum amplitude projection, front-view maximum amplitude projection of the RISP result. (d) Finger's top-view maximum amplitude projection, front-view maximum amplitude projection of the bm4d result. (e) Reconstruction result of UBP algorithm in 3D-view. (f) Reconstruction result after optimization of RISP algorithm in 3D-view. (g) Reconstruction result after optimization of bm4d algorithm in 3D-view.}
    \label{fig:figure4}
\end{figure}

\begin{figure}[htbp]
\centering

\begin{minipage}[t]{1\textwidth} 
\centering
\begin{subfigure}[t]{\textwidth}
\centering

\includegraphics[width=0.8\textwidth]{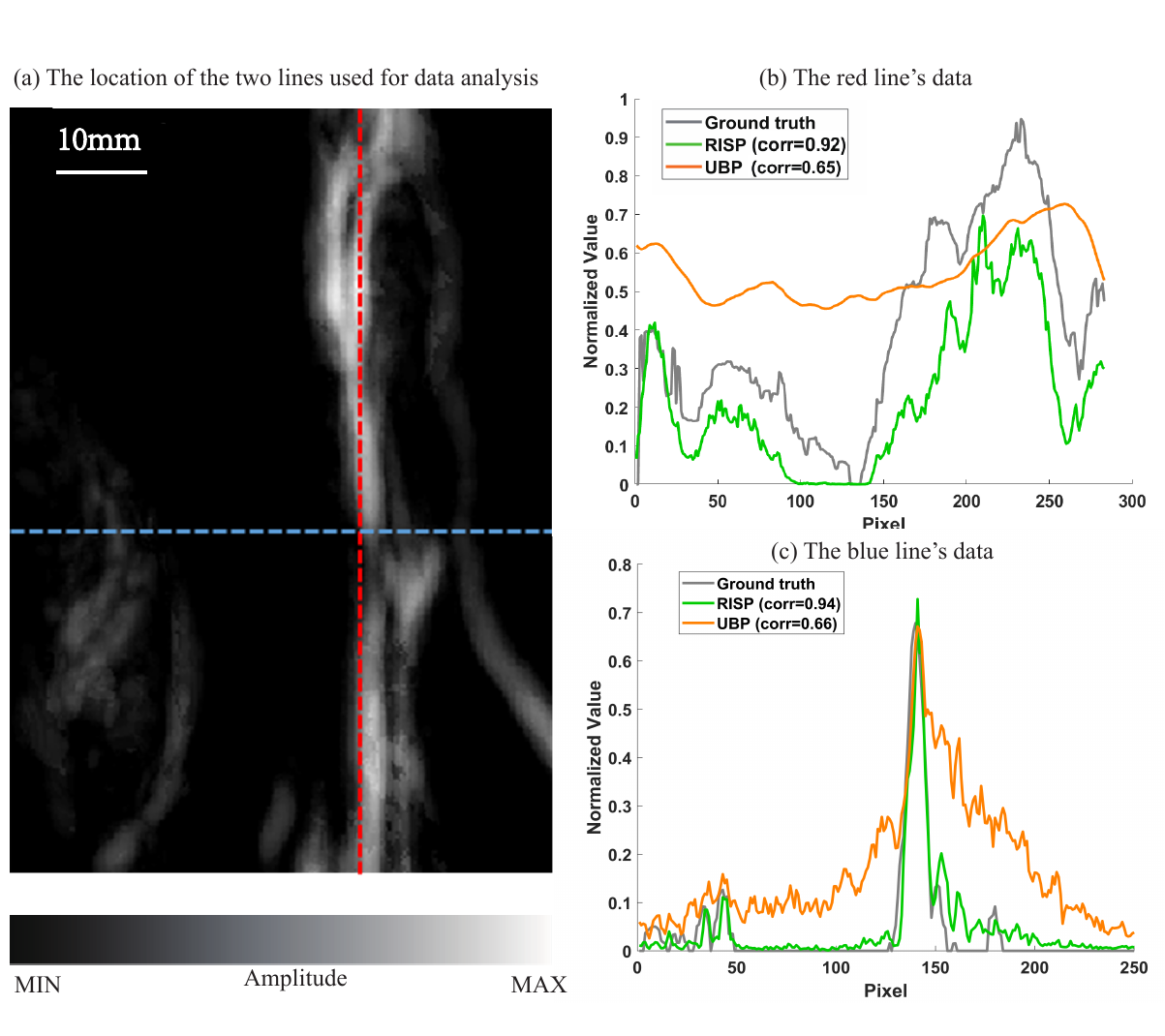} 
\label{fig5:a}
\end{subfigure}
\end{minipage}

\caption{(a) Red line with the finger information and blue line without the finger information. (b) shows the intensity distribution in red line. (c) shows the intensity distribution in blue line.}
\label{fig:Figure5}
\end{figure}

\hspace*{1em}In this example, we use the 3D PA result reconstructed by the UBP algorithm as the result optimized by the RISP algorithm. The distributions of the acoustic source are presented in Figure \ref{fig:figure4}(a), while the reconstruction results are depicted in Figure \ref{fig:figure4}(b-c). For each set of images, the upper and lower subimages correspond to the top-view maximum amplitude projection (MAP) and the front-view MAP of the 3D reconstruction, respectively. In Figure \ref{fig:figure4}(b), the reconstruction results obtained via the UBP algorithm display significant artifacts on the right side of the index finger. These arise because the laser illumination originates from the right side of the index finger, leading to a lack of information from that particular angle and, consequently, severe artifacts. When compared to the ground truth, a substantial discrepancy is evident. As shown in \ref{fig:figure4}(c), the optimization performed by the RISP algorithm significantly reduces the artifacts in the UBP results, bringing them closer to the ground truth.  The artifacts caused by incomplete angular coverage are largely corrected, and additional spurious artifacts in other regions are also mitigated, such as those observed in the middle of the finger’s top-view MAP and in the upper and lower areas of the finger’s front-view MAP. Furthermore, in the 3D view (Figure \ref{fig:figure4}(e-f)), the improvements in the PA reconstruction results after applying the RISP algorithm are clearly observable.
\\
\hspace*{2em}Specifically, we extracted data along two lines. The voxel intensity values along the lines for the three datasets are presented in Figure \ref{fig:Figure5}. In Figure \ref{fig:Figure5}(b), it is evident that, in regions containing finger information, the intensity curve of the RISP-optimized result aligns closely with the ground truth, demonstrating superior consistency compared to the UBP-only reconstruction. The correlations before and after using RISP are $0.65$ and $0.92$. In Figure \ref{fig:Figure5}(c), it can be seen that in regions lacking finger data, the noise and artifact levels in the RISP-optimized results are significantly closer to zero, while the noise and artifact levels in the UBP-only reconstruction are considerably higher. The correlations before and after using RISP are $0.66$ and $0.94$. We performed a comprehensive analysis of the peak signal-to-noise ratio (PSNR) of the reconstruction results before and after using RISP to optimize (All three data sets are normalized) and using bm4d. The results are as follows. It further demonstrates the effectiveness of the RISP optimization.

\begin{table}[h!]
\centering
\caption{The PSNR of different results}
\begin{tabular}{c|c|c|c}
\hline
     & UBP & RISP & bm4d  \\ \hline
PSNR & 4.6875dB  & \textbf{23.3810dB}  & 17.0552dB  \\ \hline
\end{tabular}
\label{tab:psnr}

\end{table}

\hspace*{2em}In conclusion, in the simulation experiments, the reconstruction results obtained using the UBP algorithm and further optimized by the RISP algorithm demonstrated excellent performance, particularly in artifact suppression, noise reduction, and the preservation and representation of true structural information.
\paragraph{Parameters Selection and Computing Environment\\}
\hspace*{1em}In this simulation experiment, the specific values of the parameters we used are as follows:
\begin{table}[h!]
\centering
\caption{3D PA image reconstruction under 1024 hemispherical system}
\begin{tabular}{{c|c}}
\hline
\textbf{Parameters} & \textbf{Value} \\ \hline
$N$ & 1024 \\ \hline
$M$ & 4096 \\ \hline
$x$ & 50 \\ \hline
$k$ & 50 \\ \hline
$\lambda_{con}$  & 0.10 \\ \hline
$\lambda_{reg}$  & 0.90 \\ \hline
$lr$  & 0.001 \\ \hline
$num\_iters$  & 500 \\ \hline
\end{tabular}

\label{tab:param_table1}
\end{table}
\\
\hspace*{2em}In this experiment, we used a graphics card type GeForce RTX 4090 and the CPU type is AMD® Epyc 9354 32-core processor $\times$ 128. In UBP algorithm reconstruction, we use Taichi\cite{23,24} for GPU acceleration. For the reconstruction with PA signal size of 50$\times$4096 and region size of 128$\times$320$\times$256, the time consuming is about 0.5s. The computation of the probability matrix takes about 35 seconds, and the entire iteration process takes about 10 seconds. So the entire RISP optimization process took less than a minute and 10 seconds ($0.5\times 50 + 35+10$). Of course, we have skills in the choice of parameters, and the choice of $x$ is generally about one-tenth of the total number of elements. We also found that when the initial value of lr is about 0.001, the iterative result can be quickly optimized. And the number of iterations is much lower than $num\_iters$, and about 50.

\subsubsection{2D PA Image Reconstruction under 256 Circle System}
\begin{figure}[htbp]
    \centering
    \begin{subfigure}[b]{\textwidth} 
        \centering
   
        \includegraphics[width=\textwidth]{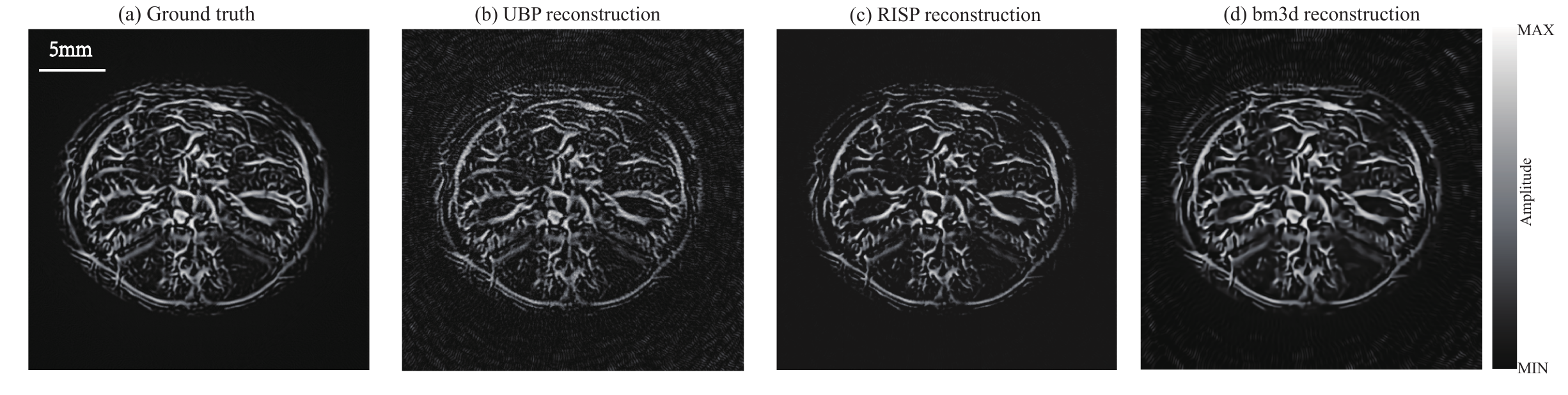} 
        \label{fig6:a}
   
    \end{subfigure}
    \caption{Comparison of 2D photoacoustic reconstruction results under 256 circle system. (a) Ground truth. (b) Reconstruction result of UBP algorithm. (c) Reconstruction result after optimization of RISP algorithm. (d) Reconstruction result after optimization of bm4d algorithm. }
    \label{fig:Figure 6}
\end{figure}

\hspace*{2em}We then used fundus blood vessels\cite{25} as target to demonstrate the high performance of the RISP algorithm (Figure \ref{fig:Figure 6}(a)). At a grid point spacing of 0.05 mm, the total grid size of the imaging area is 512$\times$512 (25.6 mm $\times$25.6 mm). The radius of the circle system is 5cm. Then we obtained PA signals collected by these 256 arrays through forward propagation of the K-Wave toolbox.
\paragraph{Comparison of Results before and after Optimization\\}
\hspace*{1em}From the Figure \ref{fig:Figure 6}, the artifacts significantly decrease after optimization by the RISP algorithm. Details of blood vessels are displayed more clearly. The PSNR results are shown in Table 3. In conclusion, in this experiment, the reconstruction results obtained using the UBP algorithm and further optimized by the RISP algorithm demonstrated excellent performance. Both the artifacts around the internal small blood vessels and the overall peripheral artifacts are weakened. And the details of the results optimized by the bm3d algorithm almost disappear.
\begin{table}[h!]
\centering
\caption{The PSNR of different results}
\begin{tabular}{c|c|c|c}
\hline
     & UBP & RISP & bm3d  \\ \hline
PSNR & 21.2393dB  &  \textbf{22.6968dB}  & 21.6542dB  \\ \hline
\end{tabular}
\label{tab:psnr2}

\end{table}

\paragraph{Parameters Selection and Computing Environment\\}
\hspace*{1em}In this simulation experiment, the specific values of the parameters we used are as follows:

\begin{table}[h!]
\centering
\caption{2D PA image reconstruction under 256 circle system}
\begin{tabular}{{c|c}}
\hline
\textbf{Parameters} & \textbf{Value} \\ \hline
$N$ & 256 \\ \hline
$M$ & 2048 \\ \hline
$x$ & 50 \\ \hline
$k$ & 100 \\ \hline
$\lambda_{con}$  & 0.10 \\ \hline
$\lambda_{reg}$  & 0.90 \\ \hline
$lr$  & 0.001 \\ \hline
$num\_iters$  & 50 \\ \hline
\end{tabular}

\label{tab:param_table1}
\end{table}
\hspace*{2em}In this experiment, the CPU type is 13th Gen Intel(R)Core(TM)i5-13500H. In UBP algorithm reconstruction, for the reconstruction with PA signal size of 50$\times$2048 and region size of 768$\times$768, the time consuming is about 0.50s. The computation of the probability matrix takes about 30 seconds, and the entire iteration process takes about 10 seconds. So, the entire RISP optimization process took about 3 minutes ($0.50\times 100 + 30+10$).

\subsection{In Vivo Human Experiment Results}
\subsubsection{3D PA image reconstruction under synthetic matrix array}
\hspace*{2em}Then in vivo human study data, including human arm was acquired by Li's work\cite{21} using synthetic matrix array. The system employed a nonfocusing linear array (customized by Imasonics, France) to receive PA signals. The linear array has 256 elements with a pitch of 0.5 mm and a kerf of 0.1 mm, i.e., a total length of 12.8 cm. The center frequency of the ultrasonic array is 3.5 MHz with over 80\% bandwidth. PA signal is amplified 1500 times via self-built two-stage amplifier, and then received by a data acquisition system (Marsonics DAQ, Tianjin Langyuan Technology Co., Ltd. China) at 40 MHz sampling rate. 
    
            
            

\hspace*{2em}In the original work, a linear array consisting of 256 elements was moved 2969 steps, with a step size of 0.1 mm for each movement, resulting in a large-scale synthetic matrix of 256 × 2969 elements. To create a sparse-view setup for better evaluation of the RISP algorithm's performance, we extracted 120 rows at equal intervals (approximately 25\% of the original data), which corresponds to a step size of 2.5 mm for each scan of the linear array. Therefore, a planar matrix array with $m \times n$ elements was used. 

\subsubsection{Comparison of Results before and after Optimization}
\begin{figure}[htbp]
    \centering
    \begin{subfigure}[b]{0.9\textwidth} 
        \centering

        \includegraphics[width=\textwidth]{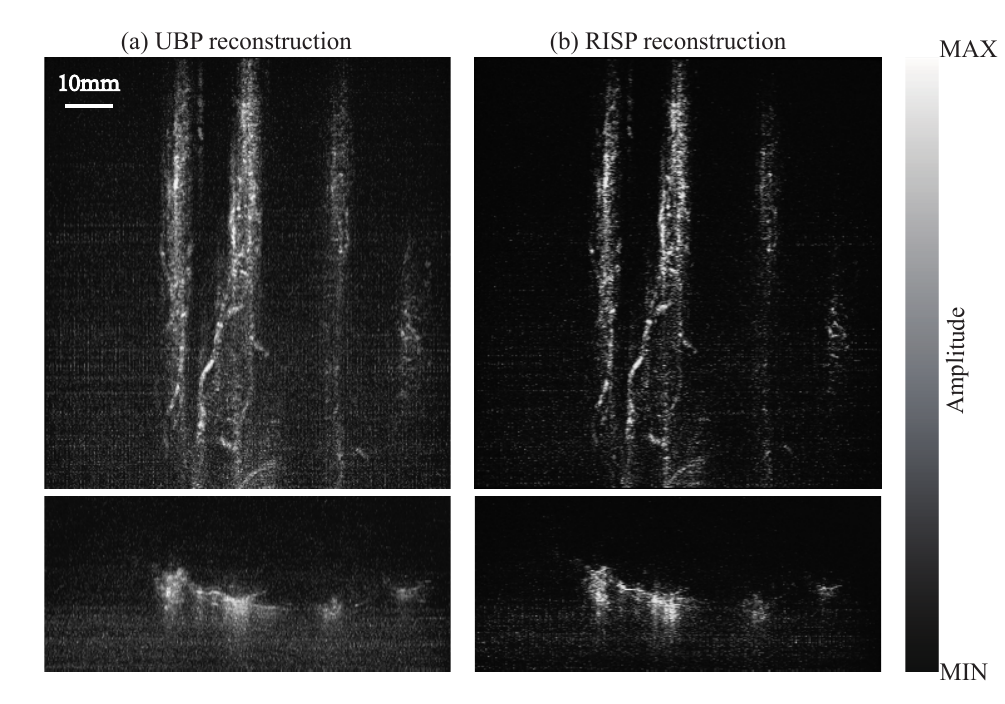} 
        \label{fig7:a}
    \end{subfigure}
   
    \caption{Comparison of 3D photoacoustic reconstruction results under synthetic matrix array. (a) Hand's top-view maximum amplitude projection, front-view maximum amplitude projection of the UBP results.\\ 
    (b) Hand's top-view maximum amplitude projection, front-view maximum amplitude projection of the RISP results. \\
    }
    \label{fig:Figure 9}
\end{figure}

\hspace*{2em}In this experiment, we utilized the RISP algorithm to optimize the reconstruction results obtained from the UBP algorithm. The detailed comparison of results before and after optimization is presented in Figure \ref{fig:Figure 9}.

\hspace*{2em}We compared the hand imaging results before and after the RISP optimization of the UBP reconstruction. From the top-view perspective, the optimization significantly reduced noise, and both thick and thin blood vessels were displayed more clearly. From the front-view perspective, in addition to making the vascular information more distinct, the lateral noise at the bottom was also effectively suppressed. And the contrast-to-noise ratio (CNR) using UBP is 8.7648, while the result after using RISP is 53.8096. These findings demonstrate that the reconstruction quality after RISP optimization is exceptionally high.

\hspace*{2em}In conclusion, in vivo human experiment, the reconstruction results obtained using the UBP algorithm and further optimized by the RISP algorithm also demonstrated excellent performance, particularly in artifact suppression, noise reduction, and the preservation and representation of true structural information.

\newpage
\subsubsection{Parameters Selection and Computing Environment}
\hspace*{2em}In this animal experiment, the specific values of the parameters we used are as follows:\\
\begin{table}[ht!]
\centering
\caption{3D PA image reconstruction under synthetic matrix array}
\begin{tabular}{{c|c}}
\hline
\textbf{Parameters} & \textbf{Value} \\ \hline
$N$ & 30720 \\ \hline
$M$ & 2048 \\ \hline
$x$ & 4000 \\ \hline
$k$ & 50 \\ \hline
$\lambda_{con}$  & 0.10 \\ \hline
$\lambda_{reg}$  & 0.90 \\ \hline
$lr$  & 0.001 \\ \hline
$num\_iters$  & 500\\ \hline
\end{tabular}

\label{tab:param_table2}
\end{table}

\hspace*{2em}In this experiment, we used a graphics card type GeForce RTX 4090 and the CPU type is AMD® Epyc 9354 32-core processor $\times$ 128. In UBP algorithm reconstruction, we use Taichi for GPU acceleration. For the reconstruction with PA signal size of 30720$\times$2048 and region size of 625$\times$600$\times$250, the time consuming is about 10.2s. For the reconstruction with PA signal size of 4000$\times$2048 and region size of 625$\times$600$\times$250, the time consuming is about 3.5s. The computation of the probability matrix takes about 140 seconds, and the entire iteration process takes about 90 seconds. So the entire RISP optimization process took less than 7 minutes ($3.5\times 50 + 140 +90$). 

\section{Discussion and Future}
\hspace*{2em}In this study, we focus on demonstrating the optimization effectiveness of the RISP algorithm applied to the reconstruction results of the UBP algorithm in the context of 3D imaging systems. Through a series of specific experiments, the RISP algorithm has shown exceptional performance in both simulated studies and live small animal experiments. Notably, from a theoretical perspective, the RISP algorithm is not constrained to a specific imaging dimension, making it applicable to both 3D and 2D imaging scenarios. Moreover, the core concept of the RISP algorithm lies in optimizing existing imaging results rather than directly reconstructing images within the signal domain. This unique characteristic grants the algorithm broad applicability, enabling effective suppression of noise and artifacts caused by transducer-related factors (e.g., transducer distribution and other system properties). Importantly, the RISP optimization is not tied to a specific reconstruction algorithm; other algorithms can also benefit from its enhancement capabilities.

\hspace*{2em}Furthermore, the potential applications of the RISP algorithm extend beyond photoacoustic computed tomography. Theoretical analysis suggests that the algorithm could be generalized to other imaging fields, such as ultrasound imaging, computed tomography (CT), and various other medical imaging or industrial inspection scenarios. In these fields, the sampling and imaging processes of various devices may also introduce artifacts and noise that vary with the properties of the transducers. In such cases, the RISP algorithm is expected to offer significant assistance.

\hspace*{2em}Going forward, we plan to expand our research in two key directions. Firstly, we aim to further investigate the optimization capabilities of the RISP algorithm across a broader spectrum of existing reconstruction algorithms, exploring its adaptability under diverse imaging conditions, modalities, and algorithmic outputs. Secondly, we intend to extend the application of the RISP algorithm to other areas, such as ultrasound imaging, CT, and potentially other advanced medical imaging techniques like magnetic resonance imaging (MRI) and positron emission tomography (PET). These efforts would further validate its extensive practical value and generalizability.

\hspace*{2em}In the long term, we aspire to make the RISP algorithm an essential tool for a wide range of imaging systems. Through continued optimization and extension studies, this algorithm has the potential to provide comprehensive solutions for improving imaging quality, resolution, and reliability across diverse application scenarios.
\newpage



\end{document}